\documentstyle[psfig]{mn}
\def\nl{\\}

\begin{document}

\title[Tilt of the Fundamental Plane]
{The tilt of the Fundamental Plane of Early-type galaxies:
wavelength dependence.
\thanks{Based on observations taken at TIRGO (Gornergrat, Switzerland). 
TIRGO is operated by CAISMI-CNR, Arcetri, Firenze, Italy.}}

\author[M. Scodeggio et al.]
{M.~Scodeggio,$^1$ G.~Gavazzi,$^2$\thanks{Osservatorio Astronomico di Brera, 
via Brera 28, 20121, Milano, Italy}
E.~Belsole,$^2$ D.~Pierini,$^3$ A.~Boselli$^4$\\
$^1$European Southern Observatory, Karl-Schwarzschild-Str. 2, D-85748
Garching bei M{\"u}nchen, Germany\\
$^2$Universit{\`a} degli Studi di Milano, dipartimento di Fisica, via Celoria
16, 20133 Milano, Italy\\
$^3$Max-Planck-Institut f{\"u}r Kernphysik, Postfach 103980, D-69117 
Heidelberg, Germany \\
$^4$Laboratoire d' Astronomie Spatiale BP8, Traverse du Syphon, F-13376 
Marseille, France\\
}

\date{}
\maketitle


\begin{abstract}
The photometric parameters $R_e$ and $\mu_e$ of 74 early-type
(E+S0+S0a) galaxies within 2 degrees projected radius from the Coma
cluster centre are derived for the first time in the near IR H band
(1.65 $\mu$m). These are used, coupled with measurements of the
central velocity dispersion $\sigma$ found in the literature, to
determine the H band Fundamental Plane (FP) relation of this cluster:
log $R_e~\propto~ A ~ log ~\sigma + b ~\mu_e$.  The same procedure is
applied to previously available photometric data in the B, V, r, I,
and K bands, to perform a multi-wavelength study of the FP relation.
Because systematic uncertainties in the value of the FP parameters are
introduced both by the choice of the fitting algorithm used to derive
the FP template, and by the presence of statistical biases connected
with the sample selection procedure, we emphasize the importance of
deriving the FP parameters in the six different photometric bands
using an identical fitting algorithm, and appropriate corrections to
eliminate the effects of sample incompleteness.  Once these
corrections are applied, we find that the FP b coefficient is stable
with wavelength ($\sim 0.35~\pm0.02$), while the A coefficient
increases significantly with increasing wavelength: from $\sim$1.35 to
$\sim 1.70~(\pm0.1$) from the optical to the IR, in agreement with an
earlier result presented by Pahre \& Djorgovski (1997).  Therefore the
slope of the FP relation, although changing with wavelength, never
approaches the virial theorem expectation (A=2.0) when the central
velocity dispersion only is used to build the FP. We also find that
the magnitude of the slope change can be entirely explained by the
presence of the well known relation between color and magnitude among
early-type galaxies.  We conclude that the tilt of the Fundamental
Plane is significant, and must be due to some form of broken homology
among early-type galaxies, while its wavelength dependence derives
from whatever mechanism (currently the preferred one is the existence
of a mass-metal content sequence) produces the color-magnitude
relation in those galaxies.
\end{abstract}

\begin{keywords}
Galaxies: elliptical and lenticular, cD -- Galaxies: fundamental
parameters -- Galaxies: stellar content -- Galaxies: clusters:
individual: Coma.
\end{keywords}

\section {Introduction}

Elliptical galaxies populate a two-dimensional manifold in the space
of their observable properties. This is generally defined in terms of
the galaxy effective radius $r_e$ (the radius that contains half of
the galaxy total light), effective surface brightness $\mu_e$ (mean
surface brightness within the effective radius) and central velocity
dispersion $\sigma$, and has been named the Fundamental Plane (FP)
(Djorgovski \& Davis 1987, Dressler et al. 1987). Alternative
representations of this manifold are also possible, and were obtained
substituting $\sigma$ with broad-band colors (de Carvalho \&
Djorgovski 1989), or relative luminosities (Scodeggio et al. 1997b).
The very existence of the FP has profound implications for the
processes that led to the formation and evolution of elliptical
galaxies (see for example Guzm{\`a}n et al.  1993, Pahre \& Djorgovski
1997, Graham \& Colless 1997a; Burstein et al. 1997).  In particular
its tightness provides a very strong constraint on the evolutionary
history of those systems. At the same time, this tightness makes the
FP a powerful tool for deriving redshift independent distance
estimates, that can be obtained with uncertainties of about 15-20 per
cent for a single galaxy.

If elliptical galaxies had perfectly homologous structural and
kinematic properties (their kinematic and density profiles were
identical), then the Virial theorem would predict the FP to have the
form $r_e \sim ~ \sigma^{2} \ast \Sigma_e^{-1}~(M/L)^{-1}$, where M/L
is the galaxy mass to light ratio. Under the assumption of a constant
M/L, the observed FP should then be log $r_e \sim ~ 2 ~ log ~\sigma +
0.4 ~\mu_e$.  In practice, however, the log $\sigma$ and $\mu_e$
coefficients have been determined to lie in the range 1.2--1.6, and
0.30--0.35, respectively. The difference between the expectation and
the observed parameters of the FP is generally named `tilt' of the
FP. This tilt is usually viewed as the product of systematic
variations of M/L along the FP, that can be parameterized as M/L $\sim
~ L^{\beta}$. The parameter $\beta$ is connected to the log $\sigma$
coefficient A by the relation $\beta$ = (2-A)/(2+A) (Pahre et
al. 1995).  Among the possible explanations that have been considered
for the tilt, there are variations in the stellar population along the
FP, due to possible variations in age, metallicity, or initial mass
function (Djorgovski \& Santiago 1993, Renzini \& Ciotti 1993, Pahre
et al. 1995, Pahre \& Djorgovski 1997), changes in the dark matter
content (Renzini \& Ciotti 1993), and deviations from homology, both
in the kinematical structure (Capelato et al. 1995) and in the matter
(and light) distribution (Hjorth \& Madsen 1995, Ciotti et al. 1996,
Graham \& Colless 1997a).  Observationally, the possibility of
distinguishing between changes in the stellar population and broken
homology is linked to the wavelength dependence of the tilt produced
by the stellar population effects, compared with a tilt that should be
strictly independent of wavelength if produced by broken homology.

Claims of a detection of systematic changes of the tilt with
wavelength have been presented, but the situation is still
uncertain. A quick and certainly incomplete survey of the FP
parameters derived by different authors during this last decade is
presented in Table 1, and illustrates well this uncertainty. The main
source of confusion about the possible wavelength dependence of the FP
tilt is in the way comparisons between several wavelengths have been
performed, that is simply comparing fits obtained by various groups,
that had used different samples, and different fitting algorithms to
derive the FP parameters. Both these differences can significantly
affect the comparison, artificially producing different FP relations
in various photometric bands.

First, systematic uncertainties introduced by the choice of a given
fitting algorithm in the derivation of the FP are most likely the
dominant source of uncertainty in the value of the FP
parameters. Different algorithms applied to the same sample can
produce values of the log~$\sigma$ coefficient, and therefore of the
tilt, that differ from each other by up to 25 per cent (the value of
the $\mu_e$ coefficient is a lot more stable, see section 5).
Therefore, since there is no well established fitting technique for FP
data-sets, it is most dangerous to compare results from different
works.  Second, sample selection biases have been generally ignored in
the derivation of FP templates. The most important bias for FP works,
that are generally based on samples of galaxies taken from rich
clusters of galaxies, is the so-called cluster population
incompleteness bias (Teerikorpi 1987, 1990; Sandage 1994a,b). Because
of the systematic lack of faint galaxies in observed samples, a biased
FP relation is obtained by simply fitting a plane to such
samples. This bias manifests itself as a flatter slope and a reduced
scatter in the observed FP, with respect to the true, unbiased
one. The amount of bias introduced in the observed FP depends on the
completeness of the sample, and therefore various FP relations,
obtained from different samples, can suffer from varying biases, and
therefore appear to have different tilt.

Although without addressing these possible problems, Pahre \&
Djorgovski (1997) were the first to derive firm evidence of a
wavelength dependence of the FP parameters. Comparing the FP relation
they obtained using K band photometric parameters with the one obtained
by Lucey et al. (1991) using V band data, they concluded that the K
band FP relation differs both from the optical one, and from the
virial expectation. Their result was reproduced by Mobasher et
al. (1997), also using K band photometric data.

With the aim of better determining the wavelength dependence of the
tilt of the FP, we present here a multi-band FP study, based on
galaxies belonging to the Coma cluster. The choice of this cluster is
guided by two main reasons: 1) the FP relation in this cluster is
known to have a small scatter (see e.g. Jorgensen et al. 1996:
Scodeggio et al. 1997a); 2) a large body of photometric and
spectroscopic data is available for this cluster. We supplement the
existing photometric data with a newly obtained set of NIR H band
(1.65 $\mu$m) observations, obtained as part of a large near-IR
imaging survey of all early type (E+S0+S0a) galaxies in the Coma
Supercluster.  Including these new data, we have now available
photometric measurements for at least 30 galaxies (that also have
spectroscopic measurements available) in 6 different bands: B, V, r,
I, H, and K.  The FP parameters in these bands are determined using a
unique fitting algorithm, therefore eliminating the most important
source of uncertainty in the differential comparisons between bands,
and the results are corrected for the effects of sample incompleteness
using a set of Monte Carlo simulations, a technique pioneered by
Giovanelli et al. (1997) for Tully-Fisher work, and applied to the FP
by Scodeggio et al. (1997a).\nl 
The remaining of this paper is organized as follows: in section 2 we
discuss the sample of galaxies used in this work; in section 3 we
present the new H band observations of 74 early-type galaxies in Coma,
in section 4 the total data-set used in this work, and in section 5
the derivation of the FP relation in the 6 different photometric
bands; the concluding discussion is presented in section 6.

\section{The Sample}

The galaxies analysed in this work are selected from the CGCG
catalogue (Zwicky et al. 1961-68) ($m_p\la15.7$), within a 2.0 degrees
radius from the X ray centre of the Coma cluster
($\alpha=12^h57^m30^s$, $\delta=28\degr15\arcmin00\arcsec$). There are
139 such objects with morphological classification E, S0, S0a. Among
these we reject those objects whose membership to the cluster is
uncertain due to either too large a velocity discrepancy for their
angular separation (see Gavazzi et al. 1995) or because they belong to
the NE filament departing from the Coma cluster (Gavazzi 1987), whose
distance could differ from that of Coma.  Among the remaining objects
we chose all those with available measurements of the central velocity
dispersion (Davies et al. 1987, Dressler 1987, Faber et al. 1989,
Lucey et al. 1991b, Scodeggio et al. 1997a), for a total of 79
objects. \nl 
For those objects we searched in the literature for
measurements of the photometric parameters used in FP studies. We have
found data for 42 galaxies in the B band (Burstein et al. 1987; Faber
et al. 1989; Saglia et al. 1993a; Prugniel \& Simien, 1996), 42 in V
(Lucey et al. 1991b), 55 in r (Jorgensen et al. 1995), 79 in I
(Scodeggio et al.  1997a, 1998a), 29 in K (Mobasher et al. 1997). To those
we add here measurements for 74 galaxies in the H band. All galaxies
are assumed to be at a distance of 96 Mpc ($\mu_o$=34.91),
derived from a mean cluster heliocentric recessional velocity of
6917 km s$^{-1}$, corrected to 7185 km s$^{-1}$ to account for the
motion with respect to the Cosmic Microwave Background (Kogut et
al. 1993), and assuming H$_o$~=~75~ km s$^{-1}$Mpc$^{-1}$.

\section{H band observations and data reduction}

The H band observations of 74 galaxies presented in this work are part
of an extensive survey of all early type galaxies with $m_{p}<15.7 ~$
in the Coma Supercluster region $11^{h}30^m<\alpha<13^{h}30^m~$;
$18\degr<\delta<32\degr$.  The entire survey will be described
elsewhere (Gavazzi et al. in preparation). The observations were
carried out with the TIRGO 1.5 m telescope at Gornergrat equipped with
the NICMOS3 camera ``ARNICA'' (Lisi et al. 1993, 1995) in 22 nights
from March 13 to April 13, 1997. The optical setting of the camera
provides a field of view of 4.3 arcmin$^2$ with 0.96 arcsec
pixels. The average seeing was 2.3 arcsec.  The observing technique
was similar to that used in previous observations of late-type
galaxies, as described in Gavazzi et al. (1996). Here we briefly
summarize some of the observing parameters relevant to this work. All
galaxies with apparent B diameter $> 1.0$ arcmin were observed with
pointing sequences which consist of eight frames centered on the
source, alternated with eight sky frames, positioned along a circular
path around the source and offset by 4 arcmin.  The on-source
positions were dithered by 10 arcsec to improve the flat-fielding, and
to facilitate the bad pixel removal. The total integration time was
6.4 min, both for the target galaxy and for the sky frames.  Galaxies
with apparent B diameter $< $ 1.0 arcmin were observed with sequences
of 9 pointings along a circular path, displaced from one-another by 1
arcmin, such that the target galaxy was always in the field. The total
integration time was 7.2 min.  To check the consistency of the
photometric calibration, 26 galaxies were observed twice or more times
during different nights. For those objects we report here the
measurements obtained on the combined frames.  The data were
calibrated with standard stars in the Elias et al. (1982) catalogue,
with a typical photometric uncertainty of 0.05 mag.

The basic image reduction was performed using standard routines in the
IRAF\footnote{IRAF (Image Reduction and Analysis Facility) is
distributed by NOAO, which is operated by the Association of
Universities for Research in Astronomy, Inc. (AURA), under cooperative
agreement with the National Science
Foundation.}-STSDAS\footnote{STSDAS(Space Telescope Science Data
Analysis System) is distributed by STScI, which is operated by AURA,
under contract to the National Aeronautics and Space
Administration.}-PROS environment.  The bias-subtracted, flat-fielded,
combined, and calibrated images were analysed using the package
GAPLHOT (developed for IRAF-STSDAS mainly by W. Freudling, J. Salzer,
and M. Haynes, and adapted by one of us (M.S.) to perform the light
decomposition of early-type galaxies).  For each frame the sky
background was determined as the mean number of counts measured in
regions of ``empty'' sky, and it was subtracted from the frame.
Sky-subtracted frames were inspected individually and the light of
unwanted superposed or nearby stars and galaxies was masked.  The
2-dimensional light distribution of each galaxy was fitted with
elliptical isophotes, using a modified version of the STSDAS ${\it
isophote}$ package.  The fit maintains as free parameters the ellipse
centre, ellipticity and position angle, and the ellipse semi-major
axis is incremented by a fixed fraction of its value at each step of
the fitting procedure. Using the fitted parameters a model of the
galaxy light distribution is obtained, which is used to compute
integrated magnitudes as a function of semi-major axis.  As an
example, we show in Fig.1 the calibrated frame of NGC 4889 prior and
after subtraction of the model fitting.

The effective radius $r_e$ and effective surface brightness $\mu_e$
(the mean surface brightness within $r_e$) of each galaxy were
obtained by fitting its radial surface brightness profile with a de
Vaucouleurs $r^{1/4}$ law (de Vaucouleurs 1948). The fit was performed
from a radius equal to twice the seeing radius, out to the outermost
isophotes for E galaxies; for S0 and S0a galaxies only the central
core was fitted. The median uncertainty on the determination of
log~$r_e$ and $\mu_e$ is 0.05 and 0.16 mag., respectively.  Total
magnitudes were obtained by extrapolating the $r^{1/4}$ fit to
infinity (E galaxies), or by extrapolating to infinity the exponential
profile that fitted the outer parts of the galaxy light profile (S0
and S0a galaxies), and adding the flux corresponding to the
extrapolated part of the profile to the one measured within the
outermost fitted galaxy isophote. The median uncertainty in the
determination of the total magnitude is 0.15 mag.

The derived parameters are listed in Table 2 as follows:\nl
Column 1: CGCG designation (CGCG field and ordinal number within that 
field). \nl
Columns 2-4: effective radius, observed and corrected for seeing 
according to the prescriptions of Saglia et al. (1993b), with 
uncertainty (in arcsec). \nl
Columns 5-6: logarithm of the effective metric (corrected) radius in kpc, 
with error. \nl
Columns 7-9: effective surface brightness, observed and corrected, with 
uncertainty (in magnitudes per square arcsec). The correction
includes the cosmological expansion (1+z)$^4$ and K-correction (taken to
be proportional to 1+z) terms, and the seeing correction, according to 
Saglia et al. (1993b). No galactic absorption correction was applied since 
A$_H$~=~0.085~A$_B$ (Pahre et al. 1996), with A$_B~\sim~0.1$ mag in the 
direction of Coma. \nl
Column 10: seeing (arcsec).

\section{Derived parameters.}

Table 3 lists the FP related parameters not already included in Table
2 for the 79 galaxies considered in this work. The table is arranged
with two lines for each galaxies. On the first line the measured
spectroscopic and photometric parameters are given, while on the
second line there are the relative measurement uncertainties. The
columns are as follows:\nl 
Columns 1-2: CGCG designation (CGCG field and ordinal number within 
that field), and NGC/IC number.\nl
Column 3: morphological type, with reference. \nl 
Column 4: Recessional velocity corrected for the bulk motion with respect 
to the Cosmic Microwave Background.\nl 
Columns 5-6 : logarithm of the central velocity dispersion with associated 
uncertainty and reference. The velocity dispersion measurements were taken 
from the following sources: a) Scodeggio et al. (1997a) (19); b) Davies 
et al. (1987) (8); c) Lucey et al. (1991b) (28); g) Faber et al. (1989) (7); 
i) Dressler (1987) (17). Although these measurements were taken using
somewhat different apertures (a used $2 \times 6$ arcsec apertures, c
used $2 \times 5.8$ arcsec ones, i lists data corrected to $2 \times
4$ arcsec ones, b and g used $4 \times 4$ arcsec ones (LCO
measurements); only one velocity dispersion in the present sample (for
the galaxy 160070) was measured in a $1.5 \times 4$ arcsec aperture
(Lick data)), they were shown to be in good relative agreement (Lucey
et al. 1991b; Scodeggio 1997), with the exception of the measurements
presented by Dressler (1987; see Davies et al. 1987, and Lucey et
al. 1991b). To compensate for the measured offset in the velocity
dispersion scale of the Dressler (1987) observations, we have applied to 
them a correction of 0.016 dex (Davies et al. 1987).
The uncertainties associated to the velocity dispersion measurements
are quoted individually in refs a and c. They are quoted 14 and 9 per cent, 
respectively, for the Lick and for the remaining observations in refs b 
and g. They are quoted 10 per cent in ref i.\nl
Columns 7-9 : B band parameters: log$~R_e$ with error, $\mu_e$ with error, 
and reference. The references are: e: Saglia et al. (1993a) (29); h: Burstein 
et al. (1987) (1); g: Faber et al. (1989) (9), m: Prugniel \& Simien, (1996) 
(3). The photometric parameters were derived by Saglia et al. (1993a) by 
fitting r$^{1/4}$ laws to the radial photometric profile derived from 
2-D data. The $\mu_e$ are corrected for cosmological effects and for 
Galactic absorption. Both $R_e$ and $\mu_e$ are corrected for seeing. 
Burstein et al. (1987) and Faber et al. (1989) derived effective circular 
diameters ($A_e$) and $\mu_e$ from a combination of photoelectric and CCD 
measurements. These data are corrected for cosmological effects and for 
Galactic absorption, but are not corrected for seeing. The uncertainties 
quoted by Saglia et al. (1993a) are 0.1 and 0.05 respectively on $\mu_e$ 
and log$~R_e$. Burstein et al. (1987) and Faber et al. (1989) quote the 
uncertainties according to four quality-classes of their observations.  The 
data from Prugniel \& Simien, (1996) are corrected for cosmological effects 
and for Galactic absorption, but are not corrected for seeing.
These authors give total magnitudes and $\mu_e$. log$~R_e$ is derived using
$m_{T}$=$\mu_e$ - 2.0 - 5log$~r_e$ \nl
Columns 10-12 : Same as cols. 6-8 for V band. Measurements are taken 
from c: Lucey et al. (1991b) (42). The photometric parameters $\mu_e$ and 
$R_e$ were derived by fitting r$^{1/4}$ laws to the radial photometric 
profile derived from 2-D data. The $\mu_e$ are corrected for cosmological 
effects, and Galactic absorption. Both $\mu_e$ and $R_e$ are corrected for 
seeing. The uncertainties on $\mu_e$ and $R_e$ are 0.14 mag and 0.028 dex 
respectively. \nl
Columns 13-15 : Same as cols. 6-8 for r band.  Measurements are taken 
from d: Jorgensen et al. (1995) (55). The photometric parameters  $\mu_e$ 
and $R_e$  were derived by fitting r$^{1/4}$ laws to the radial photometric 
profile derived from 2-D data. The $\mu_e$ are corrected for cosmological 
effects, and for Galactic absorption. Both $\mu_e$ and $R_e$ are corrected 
for seeing according to the prescriptions of Saglia et al. (1993b). The 
uncertainties on $\mu_e$ and $R_e$ are quoted by Jorgensen et al. (1995) 
for each individual galaxy. \nl
Columns 16-18 : Same as cols. 7-9 for I band. Measurements are taken 
from a: Scodeggio et al. (1997a, 1998a) (79). The photometric parameters 
$\mu_e$ and $R_e$ were derived by fitting r$^{1/4}$ laws to the radial
photometric profile derived from 2-D data. The $\mu_e$ are corrected 
for cosmological effects, and for Galactic absorption.
Both $\mu_e$ and $R_e$ are corrected for seeing according to the 
prescriptions of Saglia et al. (1993b). The uncertainties on $\mu_e$ and 
$R_e$ are given by the authors for each individual galaxy. \nl
Columns 19-21 : Same as cols. 7-9 for K band. Measurements are taken 
from Mobasher et al. (1997) (29). The photometric parameters $\mu_e$ and 
$R_e$ were derived on the integrated profiles (growth curves). Their $r_e$ 
is the radius that contains half of the asymptotic magnitude and $\mu_e$ is 
the mean surface brightness within that radius. $\mu_e$ is corrected for 
cosmological effects, and for Galactic absorption. No seeing corrections 
were applied, but the observations were carried out in 1 arcsec seeing 
conditions and the corrections were estimated $\la~0.01$ dex by the authors. 
The authors quote that the uncertainties on $\mu_e$ and log$~R_e$ are 
0.03 mag and 0.02 dex respectively.

The photometric parameters that enter in the FP determination ($\mu_e$
and $R_e$) obtained at the various wavelengths are plotted for
comparison in Fig. 2a and b against the H band values. There is a
general agreement between the various sets of measurements, except for
the K band data, and for a few galaxies marked in the figures.  The
larger scatter shown by the comparison between the H band and the B,
V, and r band data, with respect to the comparison between the H band
and the I band data, is mainly due to different fitting techniques
adopted in the case of galaxies that do not follow the r$^{1/4}$
relation at all radii (see Scodeggio et al.  1998a).  The K band data
have a markedly different distribution from those at any other band:
both $\mu_e$ and $R_e$ span half the range covered by the remaining
measurements.  When combined they produce a K band FP that extends
considerably less than in other bands (see Fig. 5), and with a slope
significantly different from the one derived by other authors at
similar wavelengths (e.g. Pahre \& Djorgovski 1997).  The reason for
this disagreement is unclear. The good correlation observed among the
parameters derived at all other bands shows that strong colours
gradients within the galaxies are to be excluded. Thus the discrepancy
is probably a spurious result, that might derive from the fact that
the method used by Mobasher et al. (1997) to determine the K band
photometric parameters differs from that used in other bands (see
above). In fact these parameters were derived from the integral growth
curves, instead of from the surface brightness profiles.  Further
support to this conclusion is given by the presence of simultaneous
discrepancies in $\mu_e$ and $r_e$: the galaxies with large $r_e$
appear to be both too bright and too small in the K band data.  Due to
the known anti-correlation between $\mu_e$ and $r_e$ (depending on the
details of the light profile, a small $r_e$ coupled with a bright
$\mu_e$ or a larger $r_e$ with a fainter $\mu_e$ produce similarly
acceptable fits), however, individual discrepancies on $r_e$ or
$\mu_e$ do not necessarily reflect in large deviations on the FP.
This is shown in Fig. 2c, where the values measured in different 
photometric bands for the combination of $\mu_e$ and $r_e$ that 
enters in the FP are compared. The good correlation between the H band
data and those in all other bands, including K, is clearly visible.

\section{The FP relation in different photometric bands}

\subsection{Fitting the FP parameters}

Finding the best-fit FP parameters is a non trivial statistical
exercise.  The fitting algorithm must minimize the `distance' of a set
of points (x,y,z) with associated uncertainties ($\epsilon$x,
$\epsilon$y, $\epsilon$z) from a plane $z=a+bx+cy~$, taking into
account that the uncertainties are not statistically independent
quantities. Here we fit the FP by minimizing the weighted sum of the
orthogonal distances of the data points from the plane. This is a
generalization to 3 dimensions of the maximum likelihood method of
Press et al. (1992) (their ''fitexy`` routine), with a modification
introduced to take into account the high degree of covariance shown by
the uncertainties on the determination of log $R_e$ and $\mu_e$ (see
Scodeggio 1997, and Scodeggio et al. 1998b for details).
Uncertainties on the FP parameters are determined using the
statistical jackknife: N sub-samples, each one composed of N--1
data-points, are extracted from the original sample of N data-points,
rejecting in turn one of the data-points.  The distribution of a
certain statistical parameter among those N sub-samples is then used
to estimate the uncertainty in the value of that same parameter,
without having to assume an a-priori statistical distribution for the
parent population of the data-set under examination (see for example
Tukey 1958, and Efron 1987).

\subsection{Completeness corrections}

The heterogeneous nature of the photometric samples used in this work
results into FP samples that have significantly different
completeness.  This fact must be taken into account, before a
meaningful comparison of the FP parameters obtained at different
photometric bands can be made.  Cluster samples used for the
determination of distance-indicator relations like the FP or the
Tully-Fisher relation suffer from an important statistical bias,
usually called cluster population incompleteness bias (Teerikorpi
1987, 1990).  Because of the systematic lack of faint cluster members
in a realistic sample, a bias is introduced in the determination of
the relation parameters: the `slope' and the scatter of the relation
are underestimated, and, as a consequence, the relation zero-point is
overestimated (see Sandage 1994a, b for a detailed analysis of this
effects). The amount of bias introduced is clearly dependent on the
degree of completeness of the sample.  Since here we are interested in
the relative values of the FP parameters in different photometric
bands, one solution to the completeness problem would be to limit our
analysis to a subset of objects that have photometric observations at
all bands, so that the sample completeness would be the same at all
bands.  This approach would however reduce significantly the size of
the available samples, and therefore significantly increase the
uncertainty in the determination of the FP parameters. For this reason
we prefer to use the full sample at all bands, and correct the FP
parameters for the effects of sample incompleteness.

The estimate of the sample completeness is obtained as follows: we
first model the B band luminosity function (LF) of (giant) early-type
galaxies as a Gaussian (Sandage et al. 1985). The peak of the Gaussian
is assumed to be at B$_{peak}$ = 16.75, which corresponds to
B$_{peak}$ = 13.0 derived by Sandage at al.  (1985) in Virgo, assuming
a differential distance modulus (Coma-Virgo) of 3.75 mag (see, for
example, the recent compilation by D'Onofrio et al. 1997). The width
of the Gaussian that better fits our total sample of CGCG galaxies is
however significantly lower than the Sandage et al. value, as shown in
Fig. 3. We find $\sigma$ = 1.1, in good agreement with Scodeggio et
al. (1997b), who determined the LF of early-type galaxies using a
composite sample of galaxies from 3 different clusters, and also in
agreement with the determination of the bright end of the Coma cluster
LF obtained by Thompson \& Gregory (1993), and by Biviano et
al. (1995).  We then assume that this LF provides a good description
of the early-type galaxies LF at all bands. This is equivalent to
assuming a constant color term to perform the transformation from B to
VrIHK, neglecting the small differences introduced by the
color-magnitude relation. This approximation is justified here, since
it does introduce only negligible effects in the computation of the FP
completeness corrections.  The solid histograms in Fig. 3 give the B
mag distribution of galaxies that enter in the FP study, binned in 0.4
mag intervals, for the samples that have available $\sigma$
measurements, and $R_e$ and $\mu_e$ measurements in the BVrIHK bands,
respectively.  The dashed-line histogram is the total CGCG sample, and
the dot-dashed line is the best fitting Gaussian LF. The ratio of the
number of galaxies in the FP sample to that expected from the Gaussian
LF in the same bin gives the sample completeness as a function of B
mag (see histograms in Fig. 3). Finally, the characteristic magnitude
B$_{compl}~$and cut-off slope are obtained fitting a Fermi--Dirac
function to the completeness distribution in the 6 observed bands
(solid line curves in Fig. 4). The Fermi-Dirac parameters so
derived are listed in Tab. 4.

More than one analytical treatment has been presented on the
derivation of sample incompleteness bias corrections for
distance-indicator relations (see Willick 1994, Sandage 1994a, b;
Sandage e al. 1995). Magnitudes however enter only indirectly into the
FP diagram, via the L-$R_e$, L-$\sigma$ and L-$\mu_e$ relations. The
systematic lack of fainter galaxies, at any given $\sigma~$ or $~
\mu_e$, reflects only marginally into a lack of small $R_e$
objects. It is thus impossible to convert directly the magnitude
completeness of a sample into a $R_e$, $\sigma~$ or $~\mu_e$
completeness.  Moreover, all the analytical treatments assume that one
is dealing with strictly magnitude limited samples, which is almost
never the case, and certainly it is not with our samples.  For these
two reasons we then use Monte Carlo simulations to derive
incompleteness corrections. This technique was pioneered by Giovanelli
et al. (1997) for Tully-Fisher work, and applied to the FP by
Scodeggio et al. (1997a).

The simulations are used to reproduce the observed relations between
the galaxy luminosity and the parameters $R_e$, $\mu_e$, and $\sigma$,
and the relation between $R_e$ and $\mu_e$ (Kormendy 1977). By then
selectively removing from the simulations the fainter galaxies, it is
possible to reproduce the effect of luminosity incompleteness on the
FP parameters (see Scodeggio 1997 for details). Because of the lack of
low luminosity objects, $R_e$ is on the average overestimated in
incomplete samples, except at the bright end of the sample. This
results in a shallower slope of the FP compared with the corrected
slope. Here the corrections were determined using an iterative
procedure, where the simulation input parameters, representing the
`true' FP relation that would be observed with a complete sample, were
adjusted until the output of the simulation, given certain
completeness parameters, reproduced the observed FP relation, as
obtained with the incomplete sample. At this point, the input
parameters were taken to be those of the correct FP relation.

\subsection{Results}

The FP relation at different wavelengths was derived using the samples
listed in Table 2 and 3, and the fitting method described in section
5.1. A 2.5 $\sigma$ clipping criterion was applied during the fitting
procedure: after a first fit was obtained, objects that deviated from
the best-fit plane by more than 2.5 times the measured dispersion were
removed from the sample, and a second fit was obtained. No second
iteration of this procedure was necessary. The resulting FP parameters
are listed in Tab. 5 together with the number of galaxies used in the
fit. An edge-on view of the FP in the various bands is given in Fig. 5
(left panels), where the solid line represents the projection of the
best-fit relation.  A clear trend in the value of the log$~ \sigma$
coefficient with wavelength is present, but this could be partly due
to the varying degree of completeness of the samples used in the fit,
with the I, and H band samples being significantly more complete than
the B, V, and r band ones. The parameters listed in Tab. 5, together
with the completeness parameters listed in Tab. 4, were used in the
Monte Carlo simulations to derive completeness-corrected FP relations
at the various wavelengths. The parameters for these corrected
relations are given in Tab. 6. Statistical uncertainties on these
parameters were obtained adding the uncertainties from the original
fits, listed in Tab. 5, to the statistical uncertainties with which
the simulations reproduce those fits.  Also, these new,
completeness-corrected FP relations are shown in the right panels of
Fig. 5 (note that the data-points plotted are still the same as in the
left panels).  The completeness corrections increase the $\sigma$
coefficients by about 0.2 on average, while the $\mu_e$ coefficients
are very little changed. The exact magnitude of the correction depends
on the sample completeness, but also on the slope of the $R_e-\mu_e$
relation.  This slope changes from approx. 0.33 in B and V, to 0.28 in
I, to approx. 0.22 in H and K, reflecting the presence of a well
defined color-magnitude relation among early-type galaxies. As a
result, samples characterized by similar completeness can have
significantly different corrections (compare, for example, the V and r
band samples).

For the K band sample we compute two different completeness
corrections. The first one is based on the results of the fit to the
incomplete sample. We have noticed, however, how the K band
photometric data span a reduced dynamic range with respect to data in
all other bands (see section 4). This might be the reason for the
somewhat discrepant $\mu_e$ coefficient that we obtain for the K band
FP relation: while in all other bands the value of this coefficient is
around 0.34, in agreement also with most of the results presented in
the literature, in K band we obtain a value of 0.40.\nl Thus we
compute a second completeness correction for the K band FP, forcing
the $\mu_e$ coefficient in the simulations to assume a value of
0.34. This results in a slightly larger correction to the log$~
\sigma$ coefficient, that becomes 1.77 instead of the 1.70 derived
from the first, normal correction. A larger K band data-set is
required to understand the origin of the discrepancy present in the
current sample.

After the application of the completeness corrections, the wavelength
dependence of the FP tilt is still visible, although the smooth
monotonic trend observed before the corrections were applied has
somewhat disappeared. If we parameterize the FP tilt in terms of a
change of the M/L~ ratio of early-type galaxies, M/L $\sim~
L^{\beta}$, as customary, we can obtain the wavelength dependence of
the parameter $\beta$, which is shown in Fig. 6. We remark that
systematic uncertainties introduced by the adopted FP fitting
technique result in an uncertainty on the absolute zero-point of the
vertical scale in this plot. The scale itself is instead a lot more
secure, due to the homogeneous treatment of the various data-sets.

\subsection{Correlation of the FP residuals}

The FP residuals in the various bands, measured as the deviation of
the individual log $R_e$ values from the best fitting FP relation, are
plotted in Fig. 7 against the H band residuals. A high degree of
correlation is clearly visible, with only few objects showing a
different behavior at different wavelengths. Since the photometric
measurements in different bands are entirely independent, the observed
correlations can hardly be explained by photometric errors. This
confirms the conclusion that photometric errors contribute a
negligible fraction of the observed FP scatter (e.g. Jorgensen et
al. 1996, Scodeggio 1997). Other measurement errors that could produce
the correlations seen in Fig. 7 include errors on $\sigma$, erroneous
assumptions on the distance of individual galaxies, or systematic
deviations from the de Vaucouleurs $r^{1/4}$ profile used in the
determination of $r_e$ and $\mu_e$.  Distance errors can be easily
excluded, as the range in log$~R_e$ residuals would correspond to a
`depth' of the sample of approximately 7000 km s$^{-1}$, comparable to
the distance to the Coma cluster itself. Systematic deviations from
the $r^{1/4}$ profile have been identified (Caon et al. 1993, Graham
et al. 1996), but the use of a more general light profile model to
derive $r_e$ and $\mu_e$ does not result into a reduced scatter in the
FP relation (Graham \& Colless 1997b). Therefore it does not appear
likely that this broken structural homology might be at the origin of
the scatter in the FP.  Velocity dispersion measurement errors are
known to provide a significant contribution to the observed FP
scatter, but this contribution is somewhat smaller than that of the
intrinsic scatter (i.e. the scatter not accounted for by measurement
uncertainties), which is estimated to be approximately 15 per cent
(e.g. Jorgensen et al. 1996, Scodeggio 1997).  If this conclusion is
correct, then differences in the intrinsic galaxy properties must be
at the origin of the FP scatter.

\section{Discussion and conclusions}

We have shown that the tilt of the FP relation changes with
wavelength, decreasing significantly as we go from the B to the K
photometric band. This trend is visible both before and after
completeness corrections are applied to the parameters of the FP, and
it is certainly not an artifact introduced by those corrections.
Their effect on the measured FP tilt is not negligible, however, and
should not be ignored when discussing the amplitude of the tilt, and
its possible origins.  On average, the result of these corrections is
to increase the log $\sigma$ coefficient in the FP by 0.2, while
leaving the $\mu_e$ coefficient unchanged. Thus a net decrease of the
tilt is produced.

The absolute value of the tilt is still somewhat uncertain, though,
because of persisting uncertainties on the fitting technique that
should be used to derive the FP parameters.  We have found differences
in the log$~\sigma$ coefficient between 0.03 and 0.22, depending on
the sample used, when using two different algorithms to fit the FP
relation (the one described in section 5.1, and the one used by
Scodeggio et al. (1997a), based on the average of the three linear
fits that can be obtained considering in turn one parameter of the FP
as the dependent variable, and the remaining two as the independent
ones). It seems however unlikely that the tilt could be made to
disappear just adopting a different fitting procedure, since the one
used here appears to produce the steepest FP template among those
presented in the literature, and still the minimum tilt observed at K
band is significantly different from zero. We therefore conclude that
a complete explanation of the tilt cannot be obtained without invoking
some contribution from broken homology or stellar population effects.
This is in agreement with previous results obtained by Pahre \&
Djorgovski (1997), and by Mobasher et al. (1997), although the net
value we obtain for the tilt at all bands is significantly smaller
than the one found in those two studies.

Two practical recipes for taking into account possible structural and
dynamical non-homology have been presented by Graham \& Colless
(1997a), and by Busarello et al. (1997). The first is based on the use
of a more general Sersic $r^{1/n}$ profile (Sersic 1968), instead of
the commonly used de Vaucouleurs $r^{1/4}$ one, to describe the light
distribution of elliptical galaxies, and on the computation of a
global velocity dispersion from the observed central one. The second
considers only dynamical non-homology, and is based on the computation
of a global `velocity', derived taking into account the total kinetic
energy (rotation plus velocity dispersion) of a galaxy. The
application of both methods to real FP data-sets resulted in a
reduction of the observed FP tilt, the most significant one obtained
when the galaxy specific kinetic energy is considered instead of the
central velocity dispersion (see Busarello et al. 1997). Since only
the latter quantity is available for most of the galaxies in our
sample, it is impossible to apply this method here with sufficient
accuracy.  However, given the magnitude of the tilt reduction
discussed by Busarello et al., it is likely that a fully corrected K
band FP relation would result to have zero tilt.  This would be in
agreement with results obtained using stellar population synthesis
models (Worthey 1994, Buzzoni 1995, Worthey, Trager \& Faber 1996)
that predict elliptical galaxies to have a constant M/L ratio at K
band.

Assuming this to be the case, the wavelength dependence of the FP tilt
finds a simple explanation in terms of one well known property of
elliptical galaxies, namely the fact that they populate a well defined
color-magnitude (CM) relation. If M/L is constant at K band, then it
must be a function of the galaxy luminosity at a different band, to
produce the CM relation. In particular, the slope of the CM relation
defines the power index of the luminosity dependence of M/L.  We have
taken the V-K vs. V CM relations of Bower et al. (1992), and of
Mobasher et al. (1997), to derive the expected M/L $\sim~ L^{\beta}$
relation in V band.  These authors find a slope of 0.082 and 0.12,
respectively, for the CM relation of Virgo and Coma E and S0
galaxies. Considering the average of these two values, we obtain an
expected relation M/L $\sim~ L_V~^{0.10}$, to compare with the results
from our FP analysis. From the tilt in the V band FP we derive M/L
$~\sim~ L_V~^{0.19}$. \nl However, in our FP analysis, the K band FP
still shows some tilt, that is equivalent to a relation M/L $\sim ~
L_V~^{0.08}$ (or M/L $\sim ~ L_V~^{0.06}$ if we were to consider the
modified completeness correction discussed in section 5.3). If we
assume that broken homology corrections erase the K band FP tilt, and
take advantage of the fact that these corrections should be
wavelength-independent, then we can derive the V band power index as
the difference between the observed V and K band indices, for a value
of 0.11 (0.13 in the case of the modified completeness corrections to
the K band FP). The good agreement between the slopes of the 
M/L $\sim ~ L_V~^{\beta}$ relation derived from the CM and the FP 
relation indicates that the wavelength dependence of the FP tilt is a
manifestation of the CM relation, seen through the effective surface
brightness instead of through total magnitudes. This should not be
surprising, given the fact that the effective radius is approximately
independent of wavelength.  Therefore systematic changes in magnitude,
at fixed $R_e$, must directly correspond to changes in $\mu_e$.  Any
mechanism that could explain the CM relation, like the presence of a
mass-metal content relation for early-type galaxies (Arimoto \& Yoshii
1987), will then explain also the presence of a wavelength dependent
tilt in the FP, without requiring extreme variations in the age or
metal content of early-type galaxies of the kind considered by Pahre
\& Djorgovski (1997).  This result is in good agreement with the
findings of Mobasher et al. (1997), obtained comparing the tilt of the
FP relation at V and K band.

An obvious future extension of the present work is along the line
traced by Burstein et al. (1997), who determined the properties of the
``generalized'' B band FP for a large number of gravitationally bound
systems, spanning from globular clusters to groups and clusters of
galaxies.  As already pointed out by these authors, the use of the B
band to compare the properties of early- and late-type galaxies has
important limitations, and one would like to view the properties of
these stellar systems in redder passbands less affected by dust and
recent star formation. With the completion of the H band survey of
early-type galaxies in the Coma Supercluster (of which the
observations reported here are a subset), we are now in the position
of accomplishing such a task, having obtained H band imaging
observations for 950 galaxies in the direction of the Coma and Virgo
clusters.  Although spectroscopic observations (including velocity
dispersion and HI or H$\alpha$ rotation measurements) are available
for little over 50 per cent of these galaxies, there will be enough
material for attempting a multi-wavelength study of the generalized FP
plane of early- and late-type galaxies.

\section*{Acknowledgments} 
We are grateful to the TAC of the TIRGO Observatory for the generous
time allocation to this project.  We thank C. Baffa, A. Borriello,
V. Calamai, B. Catinella, I. Randone, P. Ranfagni, M. Sozzi,
P. Strambio for assistance during the observations. A special thank to
V. Gavriusev for software assistance. MS wishes to thank Martha Haynes
for her assistance on using the GALPHOT package, and Riccardo
Giovanelli, Martha Haynes, Alvio Renzini, Cesare Chiosi, and Giuseppe
Longo for useful discussions.


\newpage

\begin{table*}	
\vspace{50mm}
\begin{minipage}{70mm}		

\caption{FP parameters of Coma galaxies from the literature}
\begin{tabular}{ccccccc}	
\hline
{band}    & {A}    & {B} &  {b}  & {$\beta$}  & {fit} & {ref}
\\
&    &    & {B/(-2.5)}  & {(2-A)/(2+A)}  &
\\
\hline
\\
B & 1.33 & -0.83 & 0.33 & 0.20 & LS & Dressler 1987 
\\
B & 1.39 & -0.90 & 0.36 & 0.18 & S2F & Djorgovski \& Davis 1987
\\
V & 1.23 & -0.82 & 0.33 & 0.23 & LS & Lucey et al. 1991a 
\\
r & 1.24 & -0.82 & 0.33 & 0.23 & OF & Jorgensen et al. 1996 
\\
I & 1.25 & -0.78 & 0.32 & 0.20 & AV3LS & Scodeggio et al. 1997a 
\\
I & 1.55 & -0.81 & 0.32 & 0.13 & ML & Scodeggio 1997d 
\\
k' & 1.44 & -0.79 & 0.32 & 0.16 & BLS & Pahre et al. 1995 
\\
k' & 1.66 & -0.75 & 0.30  & 0.09 & BLS & Pahre \& Djorgovski 1997 
\\
K & 1.22 & -0.76 & 0.25 & 0.24 & S3F & Mobasher et al. 1997 
\\
Virial     	     & 2 & -1.0 & 0.4 & 0.0 
\\
\hline
\end{tabular}
\end{minipage}
{} \\
LS: Least-Squares \\
S2F: Symultaneous two-variable Fit \\
OF: Orthogonal Fit \\
AV3LS: Average of 3 Least-Squares \\
ML: Maximum Likelihood \\
BLS: Bivariate Least-Squares \\
S3F: Symultaneous three-variable Fit \\
\end{table*}

\newpage

\begin{table*}
\centering
\begin{minipage}{140mm}		
\caption{H band photometric parameters}
\tiny
\begin{tabular}{lrrcrccccc}
\hline
{CGCG}    &  {$r_e$}   &  {$r_{ecorr}$} &  {$\pm$}  &  {log$R_e$}  &  {$\pm$}  
& {$\mu_e$} & {$\mu_{ecorr}$} & {$\pm$}    & {seeing}
\\
& {arcsec}   & {arcsec} &  {arcsec} & {kpc}  & {kpc} & 
{mag arcsec$^{-2}$} & {mag arcsec$^{-2}$} & {mag arcsec$^{-2}$} & {arcsec}
\\
{(1)} & {(2)} & {(3)} & {(4)} & {(5)} &
{(6)} & {(7)} & {(8)} & {(9)} & {(10)}
\\
\hline\hline

  159089 &  6.62 &  6.34 &  0.59 & 0.470 & 0.039 & 16.95 & 16.79 &  0.12 &  2.0 \\
  160013 & 32.91 & 32.87 &  1.60 & 1.185 & 0.021 & 18.79 & 18.64 &  0.05 &  2.2 \\
  160017 &  8.03 &  7.53 &  0.99 & 0.545 & 0.054 & 16.76 & 16.60 &  0.18 &  3.2 \\
  160019 &  4.00 &  3.54 &  0.36 & 0.217 & 0.039 & 15.40 & 15.17 &  0.13 &  2.2 \\
  160021 & 14.86 & 14.63 &  0.72 & 0.833 & 0.021 & 17.69 & 17.56 &  0.06 &  2.4 \\
  160022 &  9.74 &  9.38 &  0.68 & 0.640 & 0.030 & 17.14 & 17.00 &  0.09 &  2.7 \\
  160023 &  4.11 &  3.58 &  0.32 & 0.222 & 0.034 & 16.11 & 15.88 &  0.11 &  2.4 \\
  160024 & 11.26 & 10.88 &  1.34 & 0.704 & 0.052 & 17.57 & 17.41 &  0.15 &  3.0 \\
  160027 &  4.30 &  4.00 &  0.42 & 0.270 & 0.042 & 16.56 & 16.39 &  0.14 &  1.8 \\
  160028 & 10.51 & 10.28 &  0.70 & 0.680 & 0.029 & 17.01 & 16.85 &  0.09 &  2.2 \\
  160033 & 19.93 & 19.60 &  9.19 & 0.960 & 0.200 & 19.20 & 19.08 &  0.54 &  3.3 \\
  160037 &  2.93 &  2.25 &  0.36 & 0.020 & 0.053 & 15.16 & 14.78 &  0.19 &  2.4 \\
  160039 & 22.93 & 22.77 &  0.95 & 1.025 & 0.018 & 17.87 & 17.72 &  0.05 &  2.6 \\
  160042 &  6.74 &  6.32 &  0.48 & 0.469 & 0.031 & 16.48 & 16.32 &  0.10 &  2.7 \\
  160044 N &  5.80 &  5.53 &  0.21 & 0.411 & 0.016 & 16.26 & 16.10 &  0.05 &  1.9 \\
  160044 S & 13.63 & 13.48 &  0.47 & 0.798 & 0.015 & 17.13 & 17.00 &  0.04 &  1.9 \\
  160046 &  4.38 &  4.00 &  0.22 & 0.270 & 0.022 & 15.87 & 15.66 &  0.07 &  2.1 \\
  160049 &  2.39 &  1.60 &  0.54 & -0.128 & 0.097 & 15.26 & 14.70 &  0.35 &  2.9 \\
  160062 & 21.74 & 21.61 &  2.61 & 1.003 & 0.052 & 19.17 & 19.03 &  0.14 &  2.2 \\
  160063 &  4.37 &  3.66 &  0.69 & 0.231 & 0.068 & 16.08 & 15.83 &  0.23 &  2.9 \\
  160065 &  4.39 &  3.68 &  0.62 & 0.234 & 0.061 & 16.13 & 15.86 &  0.20 &  2.9 \\
  160070 &  8.88 &  8.51 &  1.05 & 0.598 & 0.051 & 17.55 & 17.38 &  0.16 &  2.7 \\
  160071 &  9.94 &  9.53 &  1.00 & 0.647 & 0.044 & 17.11 & 16.95 &  0.13 &  3.0 \\
  160079 &  8.38 &  7.75 &  1.69 & 0.557 & 0.087 & 16.90 & 16.68 &  0.28 &  3.7 \\
  160091 & 10.17 &  9.91 &  0.69 & 0.664 & 0.029 &  17.30 & 17.14 &  0.09 &  2.3 \\
  160092 &  5.20 &  4.86 &  1.39 & 0.354 & 0.116 & 17.39 & 17.22 &  0.37 &  2.1 \\
  160094 &  9.56 &  9.26 &  1.26 & 0.634 & 0.057 & 17.14 & 16.98 &  0.17 &  2.4 \\
  160097 &  4.72 &  4.30 &  0.44 & 0.301 & 0.041 & 15.95 & 15.77 &  0.13 &  2.3 \\
  160099 & 10.83 & 10.40 &  1.00 & 0.685 & 0.040 & 16.72 & 16.56 &  0.12 &  3.2 \\
  160100 &  2.25 &  1.70 &  0.24 & -0.102 & 0.045 & 15.22 & 14.81 &  0.16 &  1.9 \\
  160103 &  7.68 &  7.25 &  0.98 & 0.528 & 0.055 & 16.33 & 16.14 &  0.18 &  2.9 \\
  160105 &  7.88 &  7.65 &  0.73 & 0.552 & 0.040 & 16.57 & 16.41 &  0.13 &  1.9 \\
  160113 &  7.69 &  7.37 &  0.85 & 0.535 & 0.048 & 17.17 & 17.03 &  0.15 &  2.3 \\
  160118 & 10.80 & 10.61 &  0.84 & 0.694 & 0.034 & 16.58 & 16.47 &  0.10 &  1.9 \\
  160123 &  3.70 &  3.11 &  0.65 & 0.161 & 0.076 & 16.10 & 15.85 &  0.26 &  2.4 \\
  160124 & 32.00 & 31.84 &  2.23 & 1.171 & 0.030 & 18.24 & 18.11 &  0.08 &  2.9 \\
  160130 & 20.17 & 20.03 &  0.84 & 0.970 & 0.018 & 18.04 & 17.90 &  0.05 &  2.3 \\
  160211 &  2.85 &  1.87 &  0.34 & -0.060 & 0.051 & 15.43 & 14.87 &  0.18 &  3.7 \\
  160215 &  6.85 &  6.69 &  0.21 & 0.493 & 0.013 & 16.51 & 16.35 &  0.04 &  1.5 \\
  160216 & 17.79 & 17.52 &  2.87 & 0.911 & 0.070 & 18.65 & 18.50 &  0.19 &  2.9 \\
  160217 &  3.00 &  2.16 &  0.51 & 0.002 & 0.074 & 15.85 & 15.39 &  0.26 &  2.9 \\
  160218 &  3.68 &  2.88 &  0.60 & 0.127 & 0.071 & 15.53 & 15.18 &  0.25 &  2.9 \\
  160219 &  4.33 &  3.61 &  0.63 & 0.225 & 0.063 & 16.33 & 16.07 &  0.21 &  2.9 \\
  160220 & 13.16 & 12.85 &  3.17 & 0.777 & 0.105 & 18.54 & 18.40 &  0.30 &  2.9 \\
  160221 &  6.33 &  5.89 &  0.36 & 0.438 & 0.025 & 16.54 & 16.36 &  0.08 &  2.7 \\
  160222 &  4.98 &  4.43 &  0.81 & 0.314 & 0.071 & 16.35 & 16.17 &  0.23 &  2.7 \\
  160224 &  8.09 &  7.71 &  0.50 & 0.555 & 0.027 & 16.36 & 16.24 &  0.09 &  2.7 \\
  160227 & 11.72 & 11.54 &  0.80 & 0.730 & 0.030 & 17.84 & 17.70 &  0.08 &  1.9 \\
  160228 &  3.74 &  3.29 &  0.43 & 0.185 & 0.050 & 15.76 & 15.57 &  0.17 &  2.1 \\
  160229 & 10.15 &  9.95 &  0.80 & 0.666 & 0.034 & 17.67 & 17.55 &  0.10 &  1.9 \\
  160230 &  3.03 &  2.57 &  0.27 & 0.078 & 0.038 & 15.46 & 15.20 &  0.13 &  1.9 \\
  160231 & 20.90 & 20.79 &  1.05 & 0.986 & 0.022 & 17.83 & 17.70 &  0.06 &  1.9 \\
  160234 &  7.30 &  6.85 &  0.83 & 0.504 & 0.049 & 16.97 & 16.80 &  0.15 &  2.9 \\
  160235 &  6.06 &  5.79 &  0.76 & 0.431 & 0.054 & 17.34 & 17.16 &  0.17 &  1.9 \\
  160237 &  8.18 &  7.89 &  0.79 & 0.565 & 0.042 & 17.26 & 17.08 &  0.13 &  2.2 \\
  160238 &  6.30 &  5.96 &  0.60 & 0.443 & 0.042 & 16.53 & 16.37 &  0.13 &  2.3 \\
  160239 & 12.43 & 12.28 &  1.01 & 0.757 & 0.035 & 18.26 & 18.13 &  0.10 &  1.8 \\
  160241 & 27.66 & 27.63 &  0.66 & 1.109 & 0.010 & 17.36 & 17.24 &  0.03 &  1.8 \\
  160242 &  4.78 &  4.48 &  0.51 & 0.319 & 0.046 & 17.02 & 16.83 &  0.14 &  1.9 \\
  160244 &  2.82 &  2.47 &  0.14 & 0.061 & 0.022 & 15.44 & 15.21 &  0.07 &  1.6 \\
  160246 &  2.72 &  2.13 &  0.44 & -0.004 & 0.070 & 15.69 & 15.36 &  0.24 &  2.1 \\
  160247 & 21.84 & 21.69 &  4.56 & 1.004 & 0.091 & 19.47 & 19.39 &  0.24 &  2.4 \\
  160248 E & 10.15 &  9.87 &  3.40 & 0.662 & 0.145 & 18.06 & 17.91 &  0.40 &  2.4 \\
  160248 W &  6.02 &  5.64 &  0.52 & 0.419 & 0.037 & 16.12 & 15.95 &  0.12 &  2.4 \\
  160249 & 18.59 & 18.43 &  1.57 & 0.933 & 0.037 & 17.62 & 17.45 &  0.11 &  2.4 \\
  160250 &  6.09 &  5.79 &  0.64 & 0.431 & 0.045 & 17.21 & 17.02 &  0.14 &  2.1 \\
  160253 &  5.72 &  5.43 &  0.45 & 0.403 & 0.034 & 16.70 & 16.53 &  0.11 &  2.0 \\
  160254 & 12.22 & 11.94 &  2.52 & 0.745 & 0.090 & 18.32 & 18.18 &  0.26 &  2.6 \\
  160255 S & 7.519 &  7.15 &  1.57 & 0.522 & 0.091 & 17.88 & 17.73 &  0.28 &  2.6 \\
  160255 N &  3.87 &  3.22 &  0.81 & 0.176 & 0.091 & 15.89 & 15.64 &  0.31 &  2.6 \\
  160256 &  5.74 &  5.35 &  0.66 & 0.396 & 0.050 & 16.16 & 15.98 &  0.16 &  2.4 \\
  160258 &  8.00 &  7.67 &  0.47 & 0.553 & 0.026 & 16.74 & 16.55 &  0.08 &  2.4 \\
  160259 & 15.81 & 15.60 &  0.72 & 0.861 & 0.020 & 17.95 & 17.85 &  0.06 &  2.4 \\
  160261 & 20.33 & 20.11 &  4.61 & 0.971 & 0.098 & 18.77 & 18.64 &  0.27 &  2.9 \\
\hline\hline
\end{tabular}
\end{minipage}
\end{table*}

\normalsize

\newpage

\begin{table*}
\vbox to 220mm{\vfil
Table 3, printed in Landscape format, to go here.
\caption{}
\vfil}
\end{table*}

\newpage

\setcounter{table}{2}
\begin{table*}
\vbox to 220mm{\vfil
Table 3, printed in Landscape format, to go here.
\caption{}
\vfil}
\end{table*}

\newpage

\setcounter{table}{2}
\begin{table*}
\vbox to 220mm{\vfil
Table 3, printed in Landscape format, to go here.
\caption{}
\vfil}
\end{table*}

\newpage

\setcounter{table}{2}
\begin{table*}
\vbox to 220mm{\vfil
Table 3, printed in Landscape format, to go here.
\caption{}
\vfil}
\end{table*}

\newpage
\setcounter{page}{10}
\setcounter{table}{3}

\begin{table*}		

\begin{minipage}{50mm}		
\caption{Completeness Parameters}
\begin{tabular}{ccc}
\hline
{Band}  & {$\rm B_{compl}$}   & {cut-off}
\\
& {mag} 
\\
\hline\hline
B & 14.30 & 0.60 \\
V & 14.50 & 0.70 \\
r & 14.45 & 0.80 \\
I & 15.30 & 0.60 \\
H & 15.35 & 0.45 \\
K & 14.20 & 0.45 \\
\hline
\end{tabular}
\end{minipage}
\end{table*}

\newpage

\begin{table*}

\begin{minipage}{70mm}	
\caption{the derived FP parameters (incomplete)}
\begin{tabular}{ccccccccc}	
\hline
{Band}   & {N}  & {A}  & {$\pm$}   &
{b}  & {$\pm$}  & {dispersion} & {$\beta$} & {$\pm$}
\\
\hline\hline
B & 38 & 1.13 & 0.08 & 0.35 & 0.02 & 0.06 & 0.28 & 0.027 \\
V & 41 & 1.18 & 0.13 & 0.35 & 0.02 & 0.07 & 0.26 & 0.042 \\
r & 54 & 1.23 & 0.08 & 0.36 & 0.02 & 0.08 & 0.24 & 0.025 \\
I & 75 & 1.36 & 0.12 & 0.32 & 0.01 & 0.08 & 0.18 & 0.036 \\
H & 73 & 1.51 & 0.09 & 0.32 & 0.01 & 0.09 & 0.14 & 0.026 \\
K & 29 & 1.54 & 0.17 & 0.40 & 0.04 & 0.07 & 0.13 & 0.048 \\ 
\hline
\end{tabular}
\end{minipage}
\end{table*}
\vspace{10mm}

\begin{table*}	
\begin{minipage}{70mm}
\caption{the derived FP parameters (complete)}
\begin{tabular}{ccccccccc}
\hline
{Band}   & {N}  & {A}  & {$\pm$}   &
{b}  & {$\pm$}  & {dispersion} & {$\beta$} & {$\pm$}
\\
\hline\hline
B & 38 & 1.40 & 0.09 & 0.35 & 0.02 & 0.07 & 0.18 & 0.027 \\
V & 41 & 1.35 & 0.13 & 0.35 & 0.02 & 0.07 & 0.19 & 0.040 \\
r & 54 & 1.35 & 0.09 & 0.37 & 0.02 & 0.08 & 0.19 & 0.027 \\
I & 75 & 1.70 & 0.13 & 0.33 & 0.01 & 0.09 & 0.08 & 0.035 \\
H & 73 & 1.66 & 0.10 & 0.34 & 0.01 & 0.09 & 0.09 & 0.027 \\
K & 29 & 1.70 & 0.17 & 0.41 & 0.05 & 0.07 & 0.08 & 0.046 \\
K$_{mod}$ & 29 & 1.77 & 0.17 & 0.34 & 0.07 &  & 0.06 & 0.045 \\
\hline
\end{tabular}
\end{minipage}
\end{table*}


\setcounter{figure}{0}
\begin{figure*}
\psfig{figure=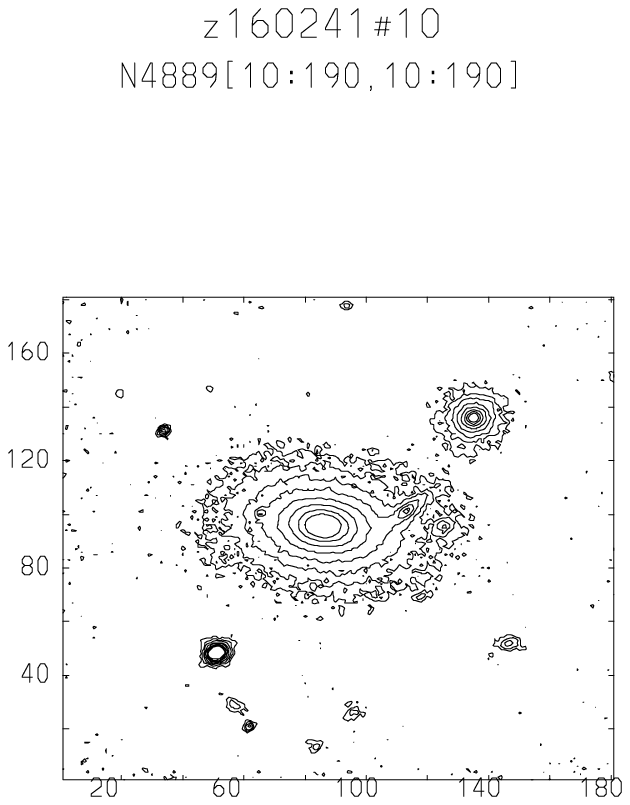}
\psfig{figure=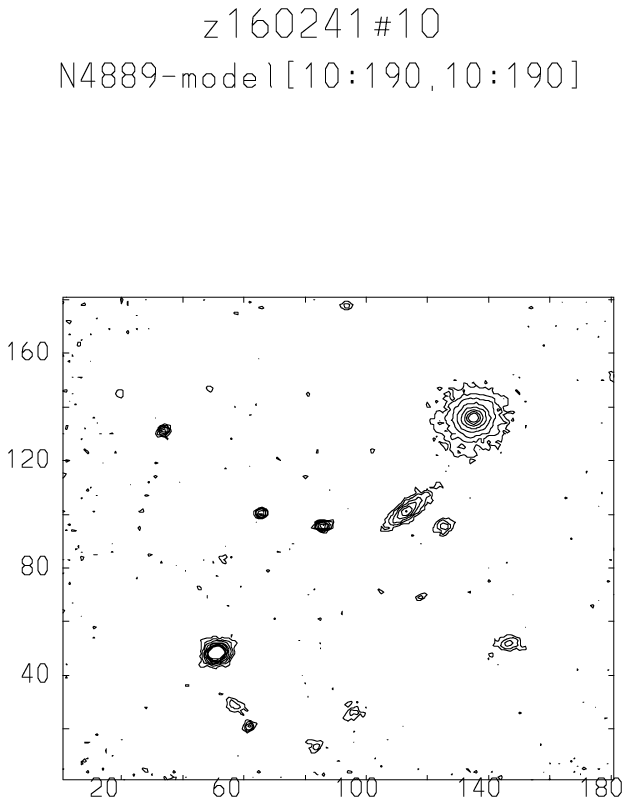}
\caption{The H band frame of NGC 4889 (160241) prior (a) and after (b)
subtraction of the model of the galaxy. The contour levels are from
19.5 to 16 $mag~arcsec^{-2}$ in steps of 0.5. North is up and East to
the left. The displayed frame is 3x3 arcmin$^2$.}
\end{figure*}

\setcounter{figure}{1}
\begin{figure*}
\psfig{figure=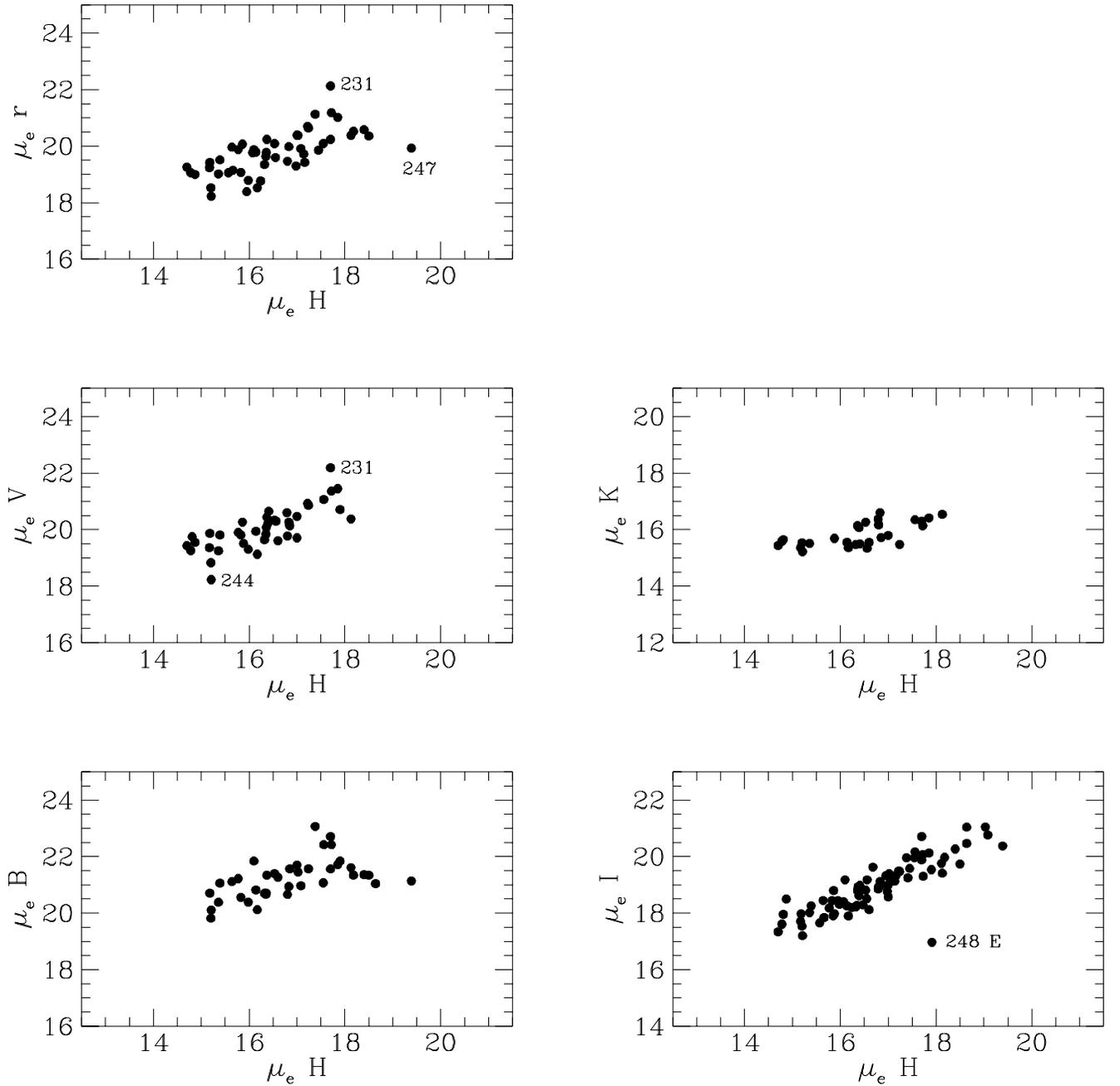}
\caption{a) The relation between $\mu_e$ as derived in the various
bands and $\mu_e$ in the H band. b) The relation between the effective
metric radius as derived in the various bands and log $R_e$ in the H
band. Few discrepant galaxies are marked, using the last digits of
their CGCG number. c) The relation between the values of the
combination of $\mu_e$ and $r_e$ that enters in the FP relation, when 
the two parameters are derived in the various bands and in the H band.}
\end{figure*}

\setcounter{figure}{1}
\begin{figure*}
\psfig{figure=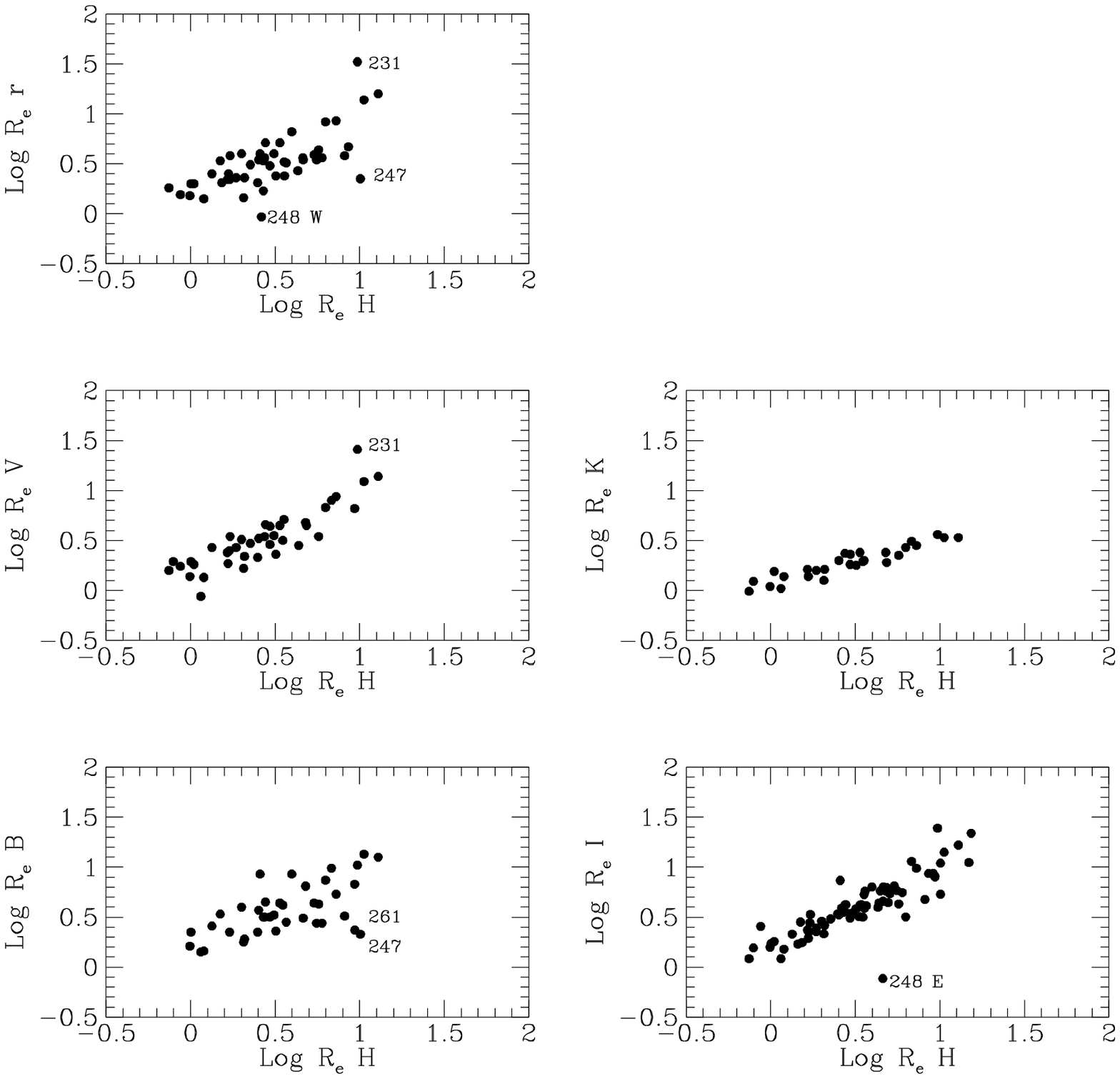}
\caption{(Continued)}
\end{figure*}

\setcounter{figure}{1}
\begin{figure*}
\psfig{figure=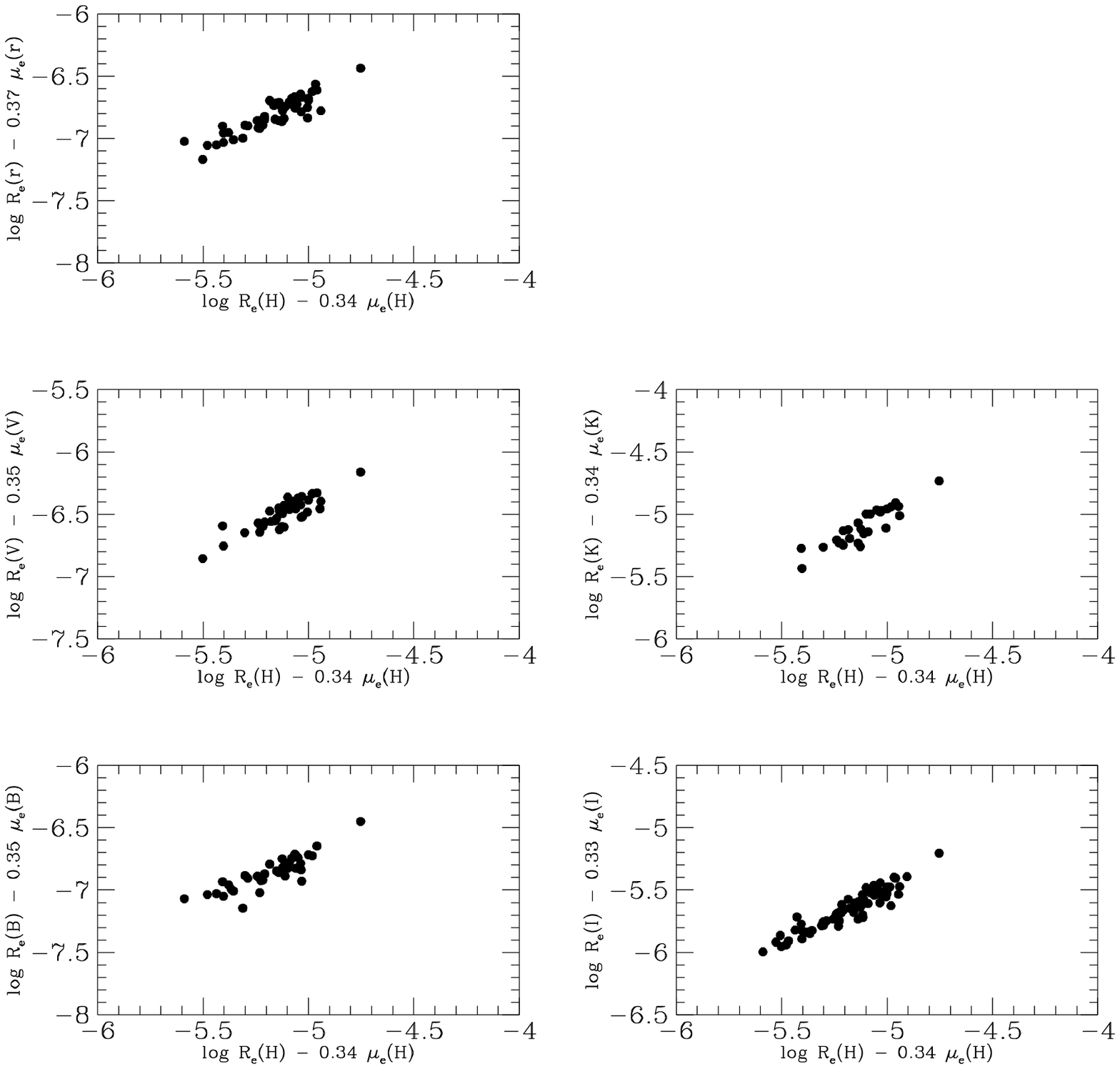}
\caption{(Continued)}
\end{figure*}

\setcounter{figure}{2}
\begin{figure*}
\psfig{figure=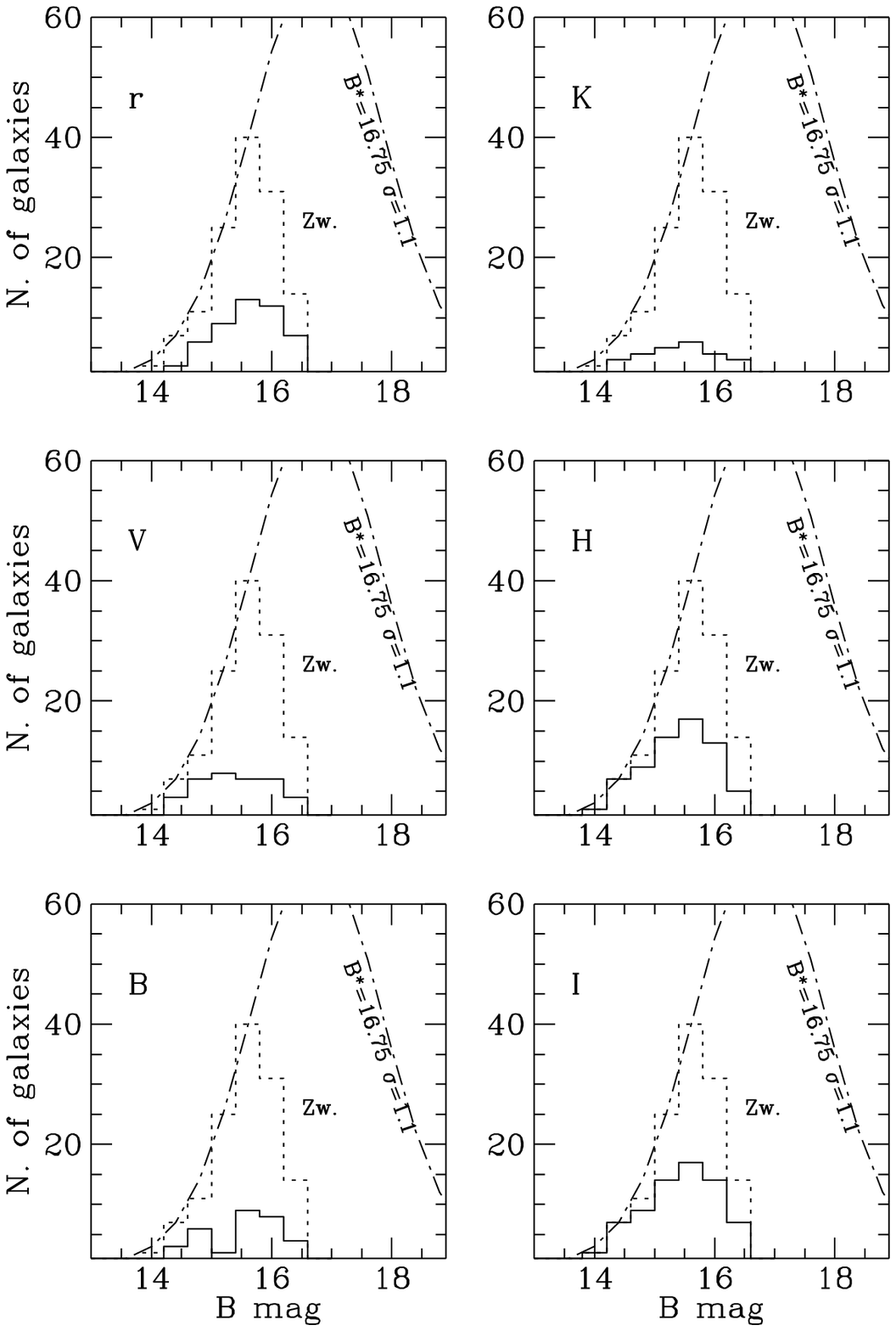}
\caption{Apparent B mag distribution of Early-type Coma cluster
galaxies in the various subsamples. The solid histograms show the
objects included in the FP study. The dotted histograms give the CGCG
distribution. The dotted-dashed curve is the adopted Gaussian
luminosity function.}
\end{figure*}

\setcounter{figure}{3}
\begin{figure*}
\psfig{figure=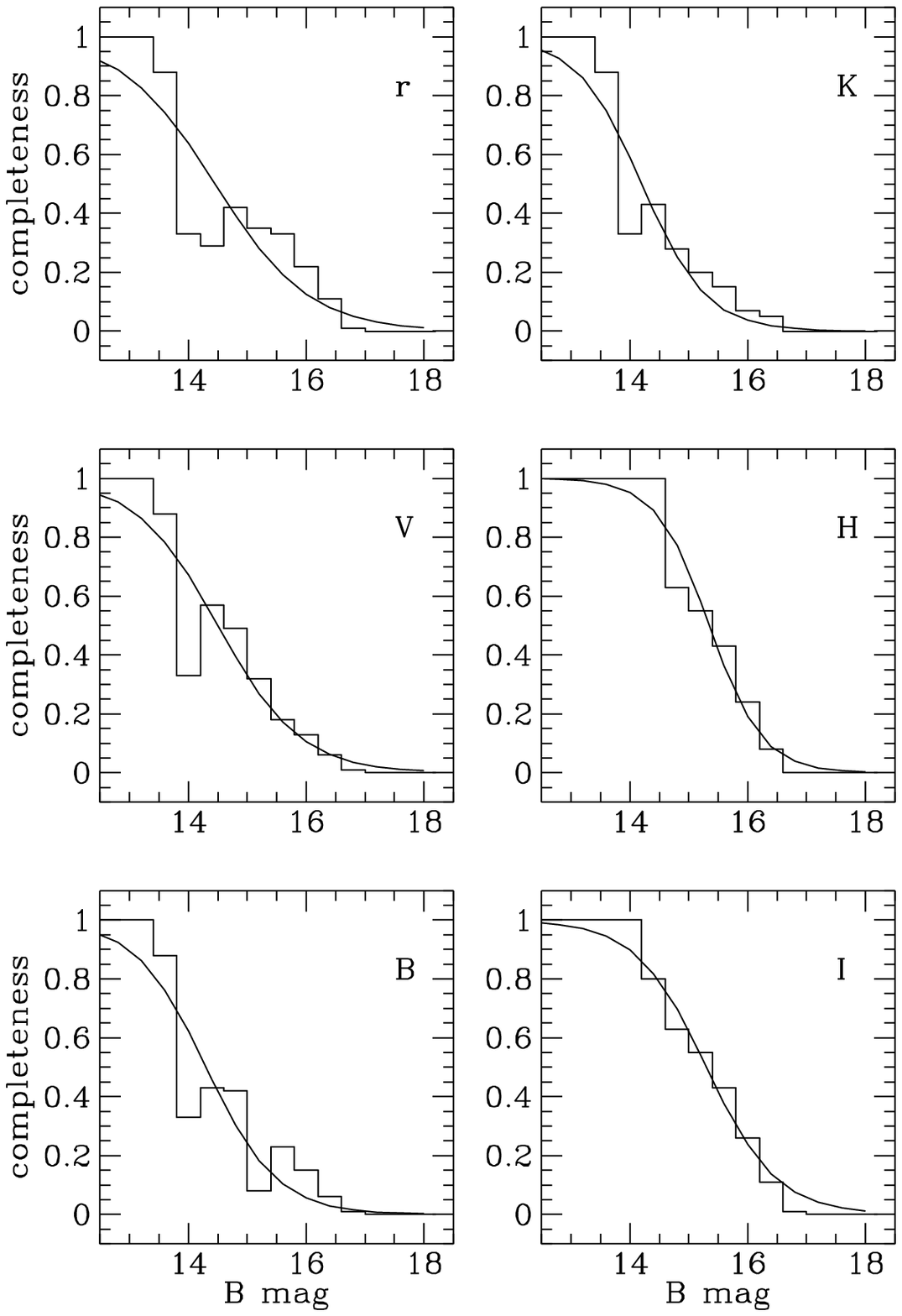}
\caption{Completeness of the FP samples in the six bands. The
histograms represent the ratio of the number of objects in each band
used in the FP determination to the number predicted by the Gaussian
luminosity function of Figure 3. The solid line is the best fit using
a Fermi--Dirac function.}
\end{figure*}

\setcounter{figure}{4}
\begin{figure*}
\psfig{figure=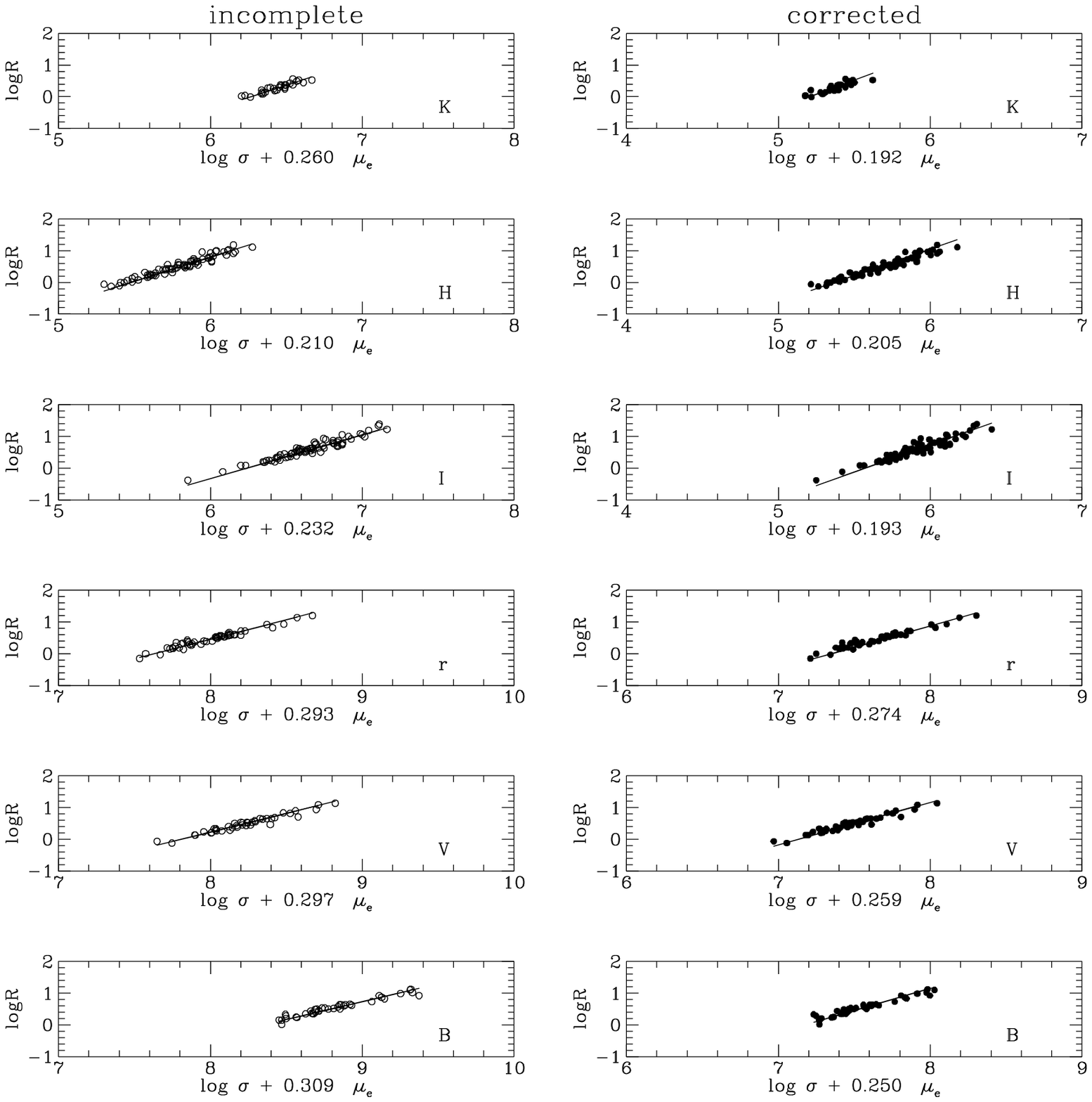}
\caption{Edge-on view of the Fundamental Plane for the six studied
bands.  Left panels: before completeness corrections. Right panels:
after completeness correction.  The solid line represents the best fit
FP parameters given in Table 5 and 6.}
\end{figure*}

\setcounter{figure}{5}
\begin{figure*}
\psfig{figure=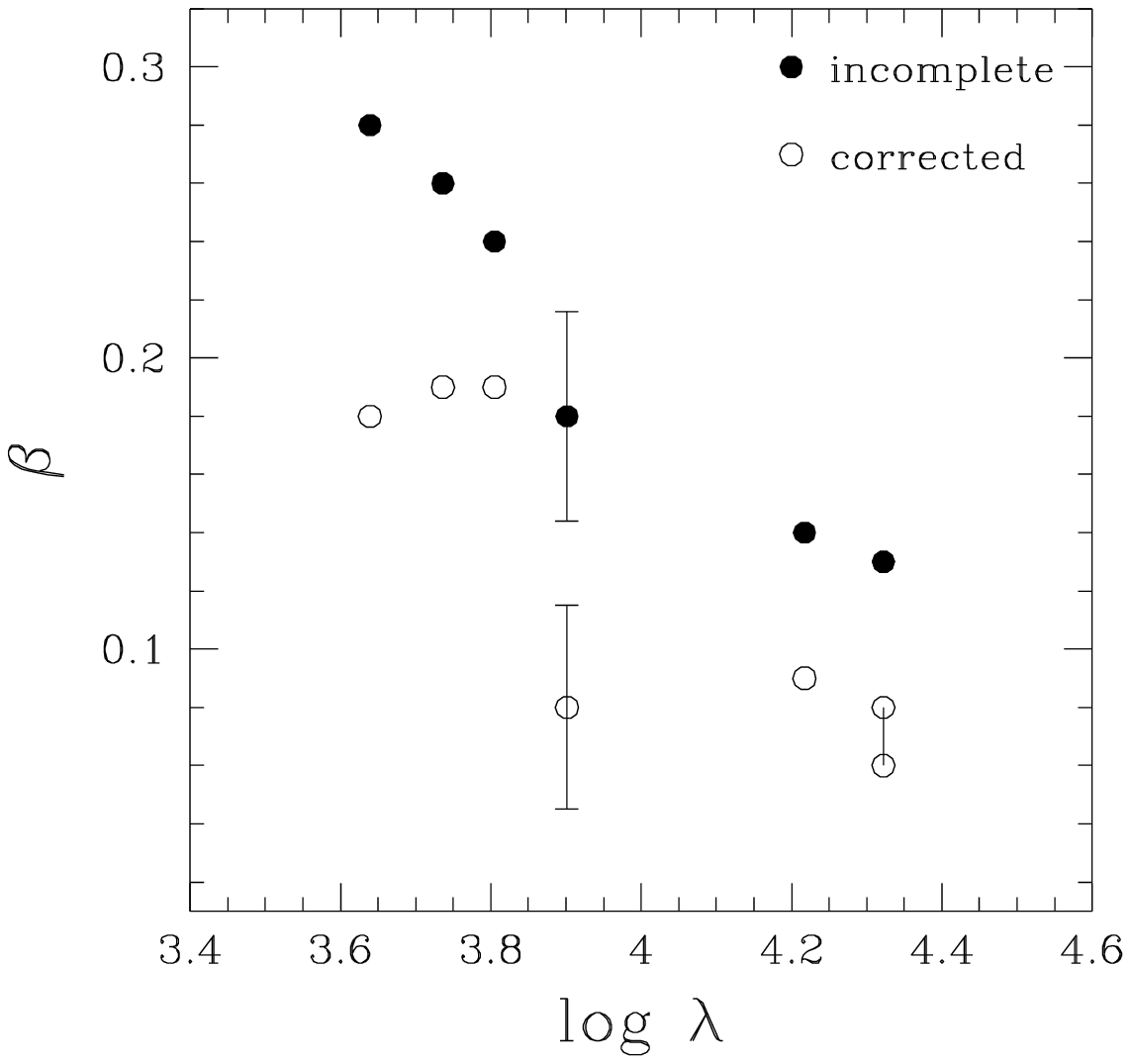}
\caption{The dependence of $\beta$ on wavelength prior and after
correction for completeness. Two values for the corrected $\beta$ at K
band are shown connected by a line (see text). Errorbars are shown
only for one point, to avoid confusion in the plot. They are however
representative of the average uncertainty with which each value of
$\beta$ is determined. Tables 5 and 6 list all individual values for
these uncertainties.}
\end{figure*}

\setcounter{figure}{6}
\begin{figure*}
\psfig{figure=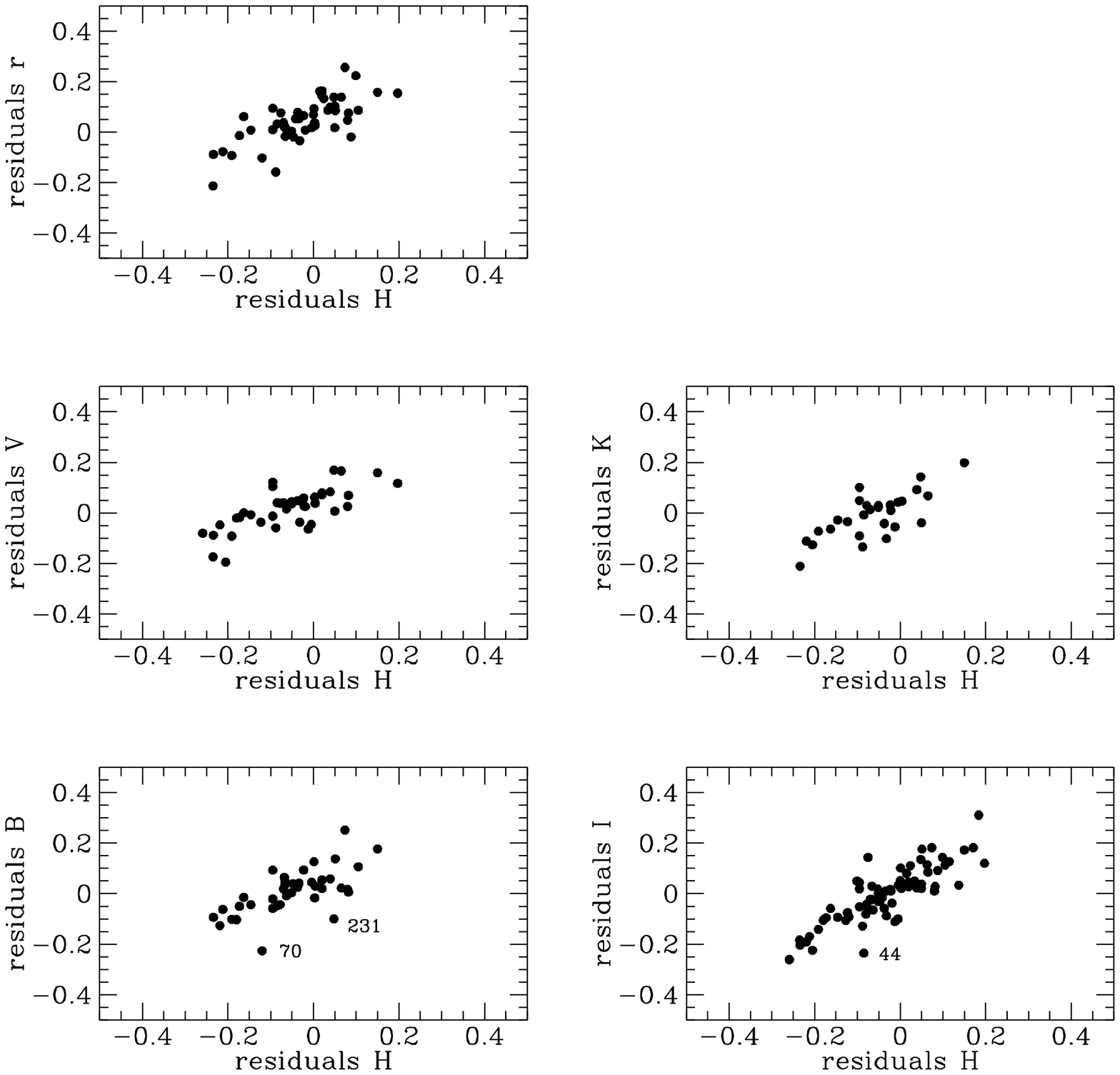}
\caption{The relation between FP residuals (measured as the difference
betwen the observed value of $\log R_e$ and the value predicted by the
FP relation) in the various bands and the H band residuals. Few
discrepant galaxies are marked, using the last digits of their CGCG
number.}
\end{figure*}

\end{document}